\pdfoutput=1
\newif\ifdraft\draftfalse

\newif\ifinlinerefs\inlinerefsfalse

\newif\ifdotikz\dotikzfalse
\dotikztrue

\documentclass{new_tlp}

\usepackage{times}
\usepackage{soul}
\usepackage{url}
\usepackage[hidelinks]{hyperref}
\usepackage[utf8]{inputenc}
\usepackage[T1]{fontenc}
\usepackage[small]{caption}
\usepackage{graphicx}
\usepackage{amsmath}
\usepackage{booktabs}
\usepackage{algorithm}
\usepackage[noend]{algorithmic}
\usepackage{ amssymb }
\newtheorem{example}{Example}
\newtheorem{theorem}{Theorem}

\usepackage[dvipsnames]{xcolor}

\newcommand\wasp{\textsc{wasp}\xspace}
\newcommand\clingo{\textsc{clingo}\xspace}

\ifdraft
\newcommand{\todo}[1]{{\Large{{\bf TODO:} \color{red}{#1}}}}
\newcommand{\comment}[1]{{\Large{\color{red}{** #1 **}}}}
\else
\newcommand{\todo}[1]{}
\newcommand{\comment}[1]{}
\fi

\usepackage{xspace}
\def\naf{\ensuremath{\raise.17ex\hbox{\ensuremath{\scriptstyle\mathtt{\sim}}}}\xspace}
\newcommand{\cP}{\ensuremath{\mathcal{P}}}
\newcommand{\cL}{\ensuremath{\mathcal{L}}}

\newcommand{\leanparagraph}[1]{\smallskip\par\noindent{\bf #1 }}

\newif\ifdotikz\dotikzfalse
\dotikztrue

\ifdotikz
\usepackage{pgfplots}
\pgfplotsset{compat=1.7}
\usetikzlibrary{plotmarks}
\pgfplotsset{
	filter discard warning=false 
	, legend cell align=left
	, minor grid style={loosely dotted, lightgray}
	, major grid style={loosely dashed, lightgray}
}

\else

\long\def\beginpgfgraphicnamed#1#2\endpgfgraphicnamed{\includegraphics{#1}}

\fi

\newenvironment{program}
{ \begin{itemize}
    \setlength{\itemsep}{0pt}
    \setlength{\parskip}{0pt}
    \setlength{\parsep}{0pt}     }
{ \end{itemize}   }

  \title[Partial Compilation of ASP Programs]
        {Partial Compilation of ASP Programs}
   \author[Cuteri et al.]
         {\!\!BERNARDO CUTERI$^1$,\! CARMINE DODARO$^1$,\! FRANCESCO RICCA\!$^1$,\! PETER SCH\"ULLER$^2$\!\!\\
         $^1$DeMaCS, University of Calabria, Italy\\
         $^2$Knowledge-based Systems Group, TU Wien, Austria \\
         \email{\{cuteri,dodaro,ricca\}@mat.unical.it, peter.schueller@tuwien.ac.at}\\
     }

\begin{document}
\label{firstpage}

\maketitle
\begin{abstract}
Answer Set Programming (ASP) is a well-known declarative formalism in logic programming. 
Efficient implementations made it possible to apply ASP in many scenarios, ranging from deductive databases applications to the solution of hard combinatorial problems.
State-of-the-art ASP systems are based on the traditional ground\&solve approach and are general-purpose implementations, i.e., they are essentially built once for any kind of input program. 
In this paper, we propose an extended architecture for ASP systems, in which parts of the input program are compiled into an ad-hoc evaluation algorithm (i.e., we obtain a specific binary for a given program), and might not be subject to the grounding step.
To this end, we identify a condition that allows the compilation of a sub-program, and present the related partial compilation technique.
Importantly, we have implemented the new approach on top of a well-known ASP solver and conducted an experimental analysis on publicly-available benchmarks. Results show that our compilation-based approach improves on the state of the art in various scenarios, including cases in which the input program is stratified or the grounding blow-up makes the evaluation unpractical with traditional ASP systems.\\
(Under consideration for acceptance in TPLP, ICLP 2019 Special Issue.)
\end{abstract}

\section{Introduction}

Answer Set Programming (ASP) is a powerful formalism that has roots in Knowledge Representation and Reasoning and is based on the stable model semantics~\cite{DBLP:journals/ngc/GelfondL91,DBLP:journals/cacm/BrewkaET11}. 
ASP is a viable solution for representing and solving many classes of problems thanks to its high expressive power and the availability of efficient systems~\cite{DBLP:conf/ijcai/GebserLMPRS18}.
Indeed, ASP has been successfully applied to several academic and industrial applications~\cite{DBLP:journals/aim/ErdemGL16} such as 
product configuration~\cite{DBLP:conf/scm/KojoMS03}, decision support systems for space shuttle flight controllers ~\cite{DBLP:conf/asp/NogueiraBGWB01}, construction of phylogenetic supertrees ~\cite{DBLP:journals/tplp/KoponenOJS15},  reconfiguration systems~\cite{DBLP:conf/cpaior/AschingerDFGJRT11}, 
and more. 
A key feature of ASP consists of the capability to model hard combinatorial problems in a declarative and compact way.
Albeit ASP is supported by efficient systems, the improvement of their performance is still an interesting research topic.

The state-of-the-art approach for solving ASP programs has two steps:
initially, variables are replaced with constants by the grounder, and the resulting equivalent variable-free program is evaluated by a propositional search-based solver computing the answer sets. 
This approach is usually referred to as the ground\&solve approach~\cite{DBLP:conf/ijcai/GebserLMPRS18}.
Moreover, ASP implementations are general-purpose, i.e., they are essentially built once for any kind of input program. 

In this paper, we propose an extended architecture for ASP systems, which allows for obtaining specific implementations for a given program and relaxes the traditional two-steps architecture by avoiding that the whole program has to be grounded upfront.

Specific implementations are obtained by introducing a technique that allows for \textit{compiling} (parts of) ASP programs into dedicated implementations.
As usual in computer science, by compilation we mean the translation of a program written in a high-level language into another programming language (usually a lower level language nearer to the machine code) to create an executable program.
To this end, we identified a condition that allows the compilation of a non-ground ASP sub-program into a C++ procedure, which simulates the behavior of that subprogram during the evaluation.
Since, in general, only parts of the input program are transformed into dedicated implementations, we name our technique \textit{partial compilation of ASP programs}.
To the best of our knowledge, \textit{this is the first attempt of compiling ASP programs in the literature}.

Whenever an entire program can be compiled an ad-hoc specialized binary is generated (this is the case for the relevant fragment of stratified normal programs); otherwise a compilable subprogram $P$ is packaged into a dynamic library that extends an existing ASP solver with an ad-hoc lazy propagator~\cite{DBLP:journals/tplp/CuteriDRS17} that simulates the behavior of $P$ during the computation of answer sets.
Note that, as it will be clearer later, compiled sub-programs are never grounded. 
One of the weak spots of the pure ground\&solve approach is that the grounding might generate a propositional program that is too big for solvers to tackle (this problem is often referred to as the grounding bottleneck) of ASP; our architecture alleviates this problem whenever the rules that are causing the bottleneck are compiled. 

An important feature of our partial compilation approach is that it can be implemented by extending in a natural way existing ASP systems that support external propagators~\cite{DBLP:conf/iclp/GebserKKOSW16,DBLP:journals/corr/abs-1811-01692}. 
This allows for keeping the benefits of existing implementations and extend their applicability and overall performance. 
In particular, our partial compilation approach has been developed by extending the state-of-the-art ASP solver \textsc{wasp}~\cite{DBLP:conf/lpnmr/AlvianoDLR15} to include propagators from dynamic libraries, and a compiler that processes a compilable sub-program and generates the corresponding source code in C++, which is finally transformed in executable code by a C++ compiler.

To assess the efficacy of our approach, we conducted an experimental analysis on publicly-available benchmarks. 
Results show that our compilation-based approach improves on the state of the art in various scenarios, including cases in which the input program is stratified or the grounding makes the evaluation less efficient with traditional ASP systems.

\section{Preliminaries}

We recall some preliminary notions that are used in the remainder of the paper.

\subsection{Answer set programming}
An ASP program $\pi$ is a finite set of rules of the form
$h_1 | \ldots | h_n \text{\tt\,:-\,} b_1, \ldots , b_m.$
where $n,m \geq 0$, $n\,{+}\,m\,{\neq}\,0$, $h_1, \ldots, h_n$ are atoms and represent the \textit{head} of the rule, while $b_1, \ldots, b_m$ are literals and represent the \textit{body} of the rule.
In particular, an \emph{atom} is an expression of the form $p(t_1, \ldots, t_k)$, where $p$ is a predicate of arity $k$ and $t_1, \ldots, t_k$ are \emph{terms}. 
Terms are alphanumeric strings and are either variables or constants.
According to Prolog conventions, only variables start with uppercase letters.
A \emph{literal} is an atom $a$ or its negation $\naf a$, where $\naf$ denotes the \emph{negation as failure}. A literal is said to be \emph{positive} if it is an atom and \emph{negative} if it is the negation of an atom.
For an atom $a$, $\overline{a} = \naf a$, for a negated atom $\naf a$, $\overline{\naf a} = a$.
A rule is called a \textit{constraint} if $n=0$, and a \textit{fact} if $n=1$ and $m=0$.

An object (atom, rule, etc.) is called \textit{ground} or \textit{propositional}, if it contains no variables. 
Given a program $\pi$, let the \textit{Herbrand Universe} $U_{\pi}$ be the set of all constants appearing in $\pi$ and the \textit{Herbrand Base}
$B_{\pi}$ be the set of all possible ground atoms which can be constructed from the predicate symbols appearing in $\pi$ with the constants of $U_{\pi}$.
Given a rule $r$, $\mathit{Ground}(r)$ denotes the set of rules obtained by applying all possible substitutions $\sigma$ from the variables in $r$ to elements of $U_{\pi}$. For a program $\pi$, the {\em ground instantiation} $\mathit{Ground}(\pi)$ of $\pi$
is the set $\bigcup_{r \in \pi} \mathit{Ground}(r)$.
Stable models of a program $\pi$ are defined using its ground
instantiation $\mathit{Ground}(\pi)$.
An interpretation $I$ for $\pi$ is a set of literals s.t. $\forall a \in B_\pi$, either $a \in I$ or $\naf a \in I$ and $l \in I \implies \overline l \notin I$.  
Given an interpretation $I$, $I^+$ denotes the set of positive literals in $I$ and $I^-$ denotes the set of negative literals in $I$.
A ground literal $l$ is {\em true} w.r.t. $I$ if $l \in I$, otherwise it is false.
An interpretation $I$ is a {\em model} for $\pi$ if, for every $r \in \mathit{Ground}(\pi)$, at
least one atom in the head of $r$ is true w.r.t. $I$ whenever all literals in the body of $r$ are true w.r.t. $I$.
The {\em reduct} of a ground program $\pi$ w.r.t. a model $I$ is the
ground program $\pi^I$, obtained from $\pi$ by (i) deleting all rules
$r \in \pi$ whose negative body is false w.r.t. $I$ and (ii)
deleting the negative body from the remaining rules.
An interpretation $I$ is a \textit{stable model} of a program $\pi$ if $I$ is a model of $\pi$, and there is no $J$ such that $J$ is a model of $\pi^I$ and $J^+ \subset I^+$ .
A program $\pi$ is \textit{coherent} if it admits at least one stable model, \textit{incoherent} otherwise.

A sub-program of $\pi$ is a set of rules $\lambda \subseteq \pi$. 
In what follows, we denote by $\cP(X)$ the set of predicate names appearing in $X$ where $X$ is an ASP expression (rule, rule head, literal, program, etc.) and we denote by $\cL(X)$ the set of literals appearing in $X$, where X is again an ASP expression. In the following, $head_r$ and  $body_r$ denote the head and the body of a rule $r$, respectively, while $body^+_r$ and $body^-_r$ denote the positive and the negative body of $r$, respectively.
Moreover, given a set of rules $\lambda$, let $\mathit{heads}(\lambda) = \{a \mid a \in head_r, r \in \lambda \}$.


\subsection{Loop unrolling and dead code elimination}\label{sec:comp}
In our work, we will mention two well-known optimizations used by compilers: \emph{loop unrolling} and \emph{dead code elimination}~\cite{DBLP:books/mk/Muchnick1997}. 
Loop unrolling is a loop transformation technique that, in the simplest formulation, removes the loop control instructions and replicates the loop body a number of times equal to the number of iterations, adjusting variables accordingly so to obtain an equivalent code. 
Dead code elimination is the removal of instructions that would never be executed, such as the body of conditional statements that are known to be false.
Such techniques are typically implemented by exploiting information that is know at compile time.

\begin{figure}[t!]
	\figrule
\begin{minipage}[c]{0.5\textwidth}
  \small
\begin{program}
	\item[] {\tt for(int j=0;j<n;j++) \{}
	\item[] \textcolor{blue}{\tt \ for(int i=0;i<3;i++) \{}
	\item[] \textcolor{blue}{\tt \ \  if(i<1) \{ a[i] = b[i] + j; \}}
	\item[] \textcolor{blue}{\tt \ \ else \ \ \ \{ b[i] = a[i] + j; \}}
	\item[] {\tt\}\}}
\end{program}
\end{minipage}
$\Longrightarrow$
\begin{minipage}[c]{0.4\textwidth}
\begin{program}
	\item[] {\tt for(int j=0;j<n;j++) \{}
	\item[] {\tt \ \  a[0] = b[0] + j;}
	\item[] {\tt \ \  b[1] = a[1] + j;}
	\item[] {\tt \ \  b[2] = a[2] + j;}
	\item[] {\tt\}}
\end{program}
\end{minipage}

\caption{Exemplification of loop unrolling and dead code elimination. The statements outlined in blue (i.e. lines 2--4) on the snippet on the left-hand side are transformed resulting in the code reported on the right-hand side.}\label{fig:compilation}
\figrule
\end{figure}

We exemplify the effect of applying loop unrolling on the snippet of C++ code reported in Figure~\ref{fig:compilation}. 
Looking at the inner {\tt for} statement (outlined in blue in Figure~\ref{fig:compilation}), we note that  the number of iterations is fixed (to 3) and is known at compile time; thus, this loop can be unrolled by a compiler by writing three instantiations of the inner block of code, one for each of the three possible values of the loop controlling variable {\tt i}, i.e., 0,1, and 2.  
In the resulting code, the three instances of the inner {\tt if}  statement (outlined in blue in Figure~\ref{fig:compilation}) contain conditions that can be evaluated at compile time (since variable {\tt i} is replaced by its actual value by loop unrolling); thus, we apply dead code elimination removing the {\tt if} statement and the code in the branch that will be never activated. 
The result of applying both loop unrolling and dead code elimination to our example is reported on the right-hand side of Figure~\ref{fig:compilation}. 
Note that the number of iterations of the outermost {\tt for} statement depends on a variable {\tt n}, thus it cannot be subject to loop unrolling at compile time because the value of  {\tt n} will be known only at execution time.

The potential benefits of applying these techniques become clear by observing that, in the original code, for each iteration of the outermost {\tt for} statement one has to perform three increments of variable {\tt i} and three evaluations of the {\tt if} statement that are not present in the equivalent transformed code. 
Loop unrolling might not always be beneficial because the program size (generally) increases, leading to potential issues such as cache misses. Nonetheless, as it will be clearer in the following, the loops that are subject to unrolling in our technique typically require very few iterations (since they are limited to the number of predicates in the program or the number of literals in rules bodies). We refer to~\cite{DBLP:books/mk/Muchnick1997} for more details about compilation techniques.

\section{Conditions for splitting and compiling}

In this section, we describe the conditions under which we allow the partial compilation.




The conditions for a sub-program to be compilable under our compilation-based approach are based on the concept of labeled dependency graph of an ASP program. 
\newtheorem{mydef}{Definition}
\begin{mydef}
Given an ASP program $\pi$, the dependency graph of $\pi$, denoted $DG_\pi$, is a labeled graph $(V,E)$ where $V$ is the set of predicate names appearing in some head of $\pi$, and $E$ is the smallest subset of $V \times V \times \{+,-\}$ such that
$(i)$ $(V_1, V_2, +) \in E$ if $\exists r \mid V_1 \in \cP(body^+_r) \wedge V_2 \in \cP(head_r)$;
$(ii)$  $(V_1, V_2, -) \in E$ if $\exists r \mid V_1 \in \cP(body^-_r) \wedge V_2 \in \cP(head_r)$; and
$(iii)$ $(V_1, V_2, -) \in E$ if $\exists r \mid V_1,V_2 \in \cP(head_r)$.
\end{mydef}
Intuitively, the dependency graph contains positive (resp., negative) arcs from positive (resp., negative) body literals to head atoms, and negative arcs between atoms in a disjunctive head.

\begin{mydef}\label{defStratified}
An ASP program $\pi$ is stratified iff $DG_\pi$ has no loop containing a negative edge. 
\end{mydef}
Definitions provided above are classical definitions for ASP programs, and
%
now we define when an ASP sub-program is compilable.
\begin{mydef}\label{defCompilable}
Given an ASP program $\pi$, an ASP sub-program $\lambda \subseteq \pi$ is \emph{compilable with respect to $\pi$} if both the following condition hold:
$(i)$ $\lambda$ is a stratified ASP program and
$(ii)$ for all $p \in \cP(\mathit{heads}(\lambda))$ it holds that $p \notin \cP(\pi \setminus \lambda)$.
\end{mydef}
Intuitively, a (sub-)program is compilable if it is stratified and does not define any predicate that appears elsewhere in the program. 
This condition often applies in practice.
Indeed, ASP encodings are often structured according to guess-and-check programming methodology,
where the checking part (typically stratified rules and constraints) is captured by the above definition.

\begin{example}
Consider the following program $\pi_1$: 
{\small
\begin{program}
\item[] {\tt (1) in(X) | out(X) :- v(X).}
\item[] {\tt (2) r(X,Y) :- e(X,Y).}
\item[] {\tt (3) r(X,Y) :- e(X,Z), r(Z,Y).}
\item[] {\tt (4) :- in(X), in(Y), not r(X,Y).}
\end{program}
}
\noindent
where \texttt{v(X)} and \texttt{e(X,Y)} model the nodes and edges of a graph, respectively.
Program $\pi_1$ contains two compilable sub-programs, one given by constraint (4) and one given by constraint~(4) together with rules (2) and (3). $\hfill \triangle$
\end{example}

Note that (sets of) constraints are always compilable; indeed, rules having no head cannot cause any cycle in the dependency graph and trivially satisfy condition (\emph{ii}) of Definition~\ref{defCompilable}.

The following result is fundamental to understand our evaluation strategy.

\begin{theorem} \label{th:compilable}
Let $\pi$ be an ASP program and  $\lambda \subseteq \pi$ be a \emph{compilable} subprogram. 
For all answer sets $M_\pi$ of $\pi$ there exists an answer set $M_{\pi \setminus \lambda}$ of $\pi \setminus \lambda$ such that $M_\pi$ is the unique answer set of the program $\{ f.~ | ~ f\in M^+_{\pi \setminus \lambda}\} \cup \lambda$.
\end{theorem}
\begin{proof}
The thesis follows from the splitting theorem~\cite{DBLP:conf/iclp/LifschitzT94}. 
Observe that the set $\cL(\pi \setminus \lambda$), i.e., the literals appearing in $\pi \setminus \lambda$, is trivially a splitting set for $\pi$, where $\lambda$ is the \emph{top program} of $\pi$ w.r.t. the splitting set, and  $\pi \setminus \lambda$ is the \emph{bottom program}.
Moreover, $\lambda$ is stratified and possibly includes constraints, thus it admits at most one answer set~\cite{DBLP:books/sp/CeriGT90}.
\end{proof}


Assuming that one can compile $\lambda$ in a specialized implementation, 
the above result suggests that one can compute an answer set $M_\pi$ of a program $\pi$ by first computing an answer set $M_{\pi \setminus \lambda}$ of $\pi \setminus \lambda$ (by using a standard ASP system), and then extending $M_{\pi \setminus \lambda}$ to $M_\pi$ by computing (resorting to the compiled implementation of $\lambda$) the answer set of the union of $\lambda$ with all atoms of $M_{\pi \setminus \lambda}$ as facts.
This sketched principle is elaborated in the following.


\begin{algorithm}[t!]
  \small
	\caption{Solving with a compiled program}\label{alg:solving}
	\begin{algorithmic}[1]
		\REQUIRE ASP program $\pi^\prime$, ASP compilable program $\lambda$
		\ENSURE An answer set of $\pi = \pi^\prime \cup \lambda$ or $\bot$ if $\pi$ is incoherent
		\STATE $\lambda^\mathit{eval}=$compile($\lambda$)
		\STATE $M_{\pi^\prime}$ = answer\_set($\pi^\prime$)
		\WHILE{$M_{\pi^\prime} \neq \bot$}
		\STATE $(C,M_{ext})$ = $\lambda^\mathit{eval}(M_{\pi^\prime})$
		\IF{$C \neq \emptyset$} \STATE{$\pi^\prime = \pi^\prime \cup C$}
		\ELSE \RETURN $M_{ext}$
		\ENDIF
		\STATE $M_{\pi^\prime}$ = answer\_set($\pi^\prime$)
		\ENDWHILE
		\RETURN $\bot$
	\end{algorithmic}
\end{algorithm}
\section{Architecture for Partial Compilation}\label{sec:architecture}
The architecture for evaluating ASP programs with partial compilation is formalized in Algorithm \ref{alg:solving}.
The algorithm takes as input two ASP programs $\pi^\prime$ and $\lambda$, where $\lambda$ is compilable with respect to $\pi \,{=}\, \pi^\prime \,{\cup}\, \lambda$, and computes one answer set of $\pi$ if it exists, otherwise it returns $\bot$ to denote that the input is incoherent. In the following  $\lambda_R$ denotes the set of stratified rules with non-empty head in $\lambda$ and $\lambda_C$ the set of constraints in $\lambda$. 
First the program $\lambda$ is compiled obtaining the procedure $\lambda^\mathit{eval}$.
Then, procedure $answer\_set$ (i.e., a standard ASP system comprising grounder and solver) is called to compute an answer set $M_{\pi^\prime}$ of $\pi^\prime$.
If $\pi^\prime$ is incoherent then $answer\_set$ returns $\bot$ and Algorithm~\ref{alg:solving} terminates returning $\bot$.
Otherwise, $M_{\pi^\prime}$ is provided as input to the compiled program $\lambda^\mathit{eval}$, which returns a pair $(C,M_{ext})$, where $C$ is a set of ground constraints having in the body only literals from $B_{\pi^\prime}$, and $M_{ext}$ is an answer set for $\pi^\prime \cup \lambda_R$.
We use subscript \emph{ext} to denote that it is the extension of the answer set of $\pi^\prime$ with the answer set of $\lambda_R$.
Importantly, $C$ models a sufficient condition for discarding $M_{\pi^\prime}$, and possibly also other candidate answer sets $M'_{\pi^\prime}$ of $\pi^\prime$ that cannot be extended to answer sets of $\pi$ because $M'_{\pi^\prime} \,{\cup}\, \lambda$ is incoherent.
If $C \,{=}\, \emptyset$ then $\lambda \,{\cup}\, M_{\pi^\prime}$ is coherent, Algorithm~\ref{alg:solving} terminates, returning $M_{ext}$ (line 8) which is an answer set of $\pi$ (by Theorem~\ref{th:compilable}).
Otherwise, if $C \neq \emptyset$, $C$ is added to $\pi^\prime$, so that the subsequent call to $answer\_set$ searches for another answer set of $\pi^\prime$.
The execution continues until $\pi^\prime$ is detected to be incoherent (line 3), and $\bot$ is returned (line 10), or an answer set is found.

The correctness of this evaluation strategy follows trivially from Theorem~\ref{th:compilable}, once we have correct algorithms for $answer\_set$, and $\lambda^\mathit{eval}$. 
How to obtain $answer\_set$ is well-known, thus in the following we describe 
the way in which we obtain $\lambda^\mathit{eval}$.

\begin{algorithm}[t]
\small
	\caption{BottomupEvaluation()}\label{alg:bottom-up}
	\begin{algorithmic}[1]
		\REQUIRE ASP program $\lambda = \lambda_R \cup \lambda_C $, an answer set $M_{\pi^\prime}$ of $\pi^\prime$
		\ENSURE  A set of ground constraints $C$ and an interpretation $M_{ext}$
		\STATE {$R= M_{\pi^\prime}$}
		\color{blue}
		\STATE {$DG = \mathrm{dependency\_graph}(\lambda)$}
		\color{black}
		\color{blue}
		\STATE {$SCCs = \mathrm{topological\_sort}(DG)$}
		\FORALL{$SCC \in SCCs$}
		\FORALL{predicate $P \in SCC$}\label{alg:loop1}
		\FORALL{exit rules $r \in \lambda_R$ with $P \in \mathcal{P}(head_r)$}\label{alg:eval1}
		\STATE $S = \mathrm{starter\_atom}(r)$\label{alg:starter}
		\color{black}
		\FORALL{$s \in R_S$}
		\STATE $R_P = R_P \cup \mathrm{evaluate}(r, s, R)$ 
		\ENDFOR
		\ENDFOR
		\ENDFOR
		\color{blue}
		\FORALL{predicate $P \in SCC$}\label{alg:loop2}
		\color{black}
		\STATE $W_P = R_P$
		\color{blue}
		\ENDFOR
		\WHILE{$\exists W_P \in W \mid W_P \neq \emptyset$}
		\WHILE{$W_P \neq \emptyset$}
		\FORALL{$r \in \lambda_R \mid \mathcal{P}(head_r) \in SCC, P \in \mathcal{P}(body^+_r)$}\label{alg:eval2}
		\color{black}
		\FORALL{$s \in W_P$}
		\STATE $E = \mathrm{evaluate}(r, s, R)$
		\STATE $W_{\mathcal{P}(head_r)} = W_{\mathcal{P}(head_r)} \cup (E \setminus R_{\mathcal{P}(head_r)})$
		\STATE $R_{\mathcal{P}(head_r)} = R_{\mathcal{P}(head_r)} \cup E$
		\STATE $W_P = W_P \setminus \{s\}$
		\ENDFOR
		\ENDFOR
		\ENDWHILE
		\ENDWHILE
		\ENDFOR
		\STATE $K = \emptyset$\label{alg:constr-start}
		\FORALL{$r \in \lambda_C$}
		\STATE $S = \mathrm{starter\_atom}(r)$\label{alg:starter-constraint}
		\color{black}
		\FORALL{$s \in R_S$}
		\STATE $K = K \cup \mathrm{ground(r, s, R)}$
		\ENDFOR
		\ENDFOR
		\color{black}
		\STATE $M_{ext} = R$
		\STATE $C = \emptyset$
		\FORALL{$c \in K$}
		\STATE $C = C \cup \{BuildConstraint(c,M_{\pi^\prime},M_{ext}, \lambda_R)\}$
		\ENDFOR
		\RETURN $(C,M_{ext})$
	\end{algorithmic}
\end{algorithm}

\section{Compilation of sub-programs}
In this section, we describe our strategy for compiling a sub-program $\lambda$ to obtain procedure $\lambda^\mathit{eval}$. 
In order to simplify the presentation, we first describe a general-purpose evaluation strategy that is valid for any compilable input program, and then we describe how this strategy can be instantiated by transforming $\lambda$ into a $\lambda$-specific algorithm that evaluates $\lambda$ w.r.t.~an answer set $M_{\pi^\prime}$ of $\pi^\prime$ by applying loop unrolling and dead code elimination (see Section~\ref{sec:comp}).
The general purpose strategy is essentially composed of two components: $(i)$ a procedure for computing bottom-up an answer set of a compilable program and a set of facts, and in case there does not exists one, $(ii)$ an algorithm computing a set of constraints that are violated by the input facts.


\paragraph{Generic Bottom-up Evaluation.}
Historically, bottom-up semi-na\"ive algorithms are the standard way to evaluate stratified programs~\cite{DBLP:books/sp/CeriGT90}.
We also adopt this algorithm, that we have refactored and exemplified in pseudo-code in
Algorithm \ref{alg:bottom-up} to make more clear how compilation specializes it depending on the program in input. 
In the algorithm, $SCCs$ denotes the topologically ordered set of the strongly connected components of the dependency graph $DG_\lambda$; and given a set of literals $X$, $X_P$ denotes the set of literals in $X$ whose predicate is $P$, thus $W_P$ and $R_P$ denotes sets of literals w.r.t. predicate $P$ and we call them the \emph{working set} and the \emph{result set} of predicate $P$, respectively.
 
The evaluation of $\lambda$ starts with the computation of the dependency graph $DG$ of $\lambda$.
Once the dependency graph is computed, the evaluation considers one strongly connected component (SCC) at a time, following a topological sort of the dependency graph. 
The for loops at line \ref{alg:loop1} and \ref{alg:loop2} iterate over all predicate names in the current SCC. 
Rules are classified into \textit{exit} and \textit{recursive}.
A rule $r$ is an exit rule for an SCC $S$ if all predicates in $\cP(body_r)$ belong to a component that precedes $S$ in the topological sort. Otherwise, $r$ is said to be recursive, i.e. there is some body predicate in the body of $r$ that belongs to $S$.
For each SCC, \textit{exit rules} are evaluated first (line~\ref{alg:eval1}), while \textit{recursive rules} are evaluated whenever all exit rules of the SCC have been evaluated (line~\ref{alg:eval2}).

Rules are evaluated as nested join loops~\cite{DBLP:books/sp/CeriGT90,DBLP:books/daglib/0020812} and the join starts with an atom, called \textit{starter atom}.
For exit rules and constraints, we have only a single join loop and the starter atom is selected among positive body atoms of the rule.
For recursive rules, we might have several join loops, and each starter atom is selected among atoms whose predicate belongs to the recursive component. The reason is that exit rules do not produce new atoms in the same component while recursive rules produce new atoms that can trigger new joins.
A nested join loop of a rule $r$ and a starter atom $s$ is implemented by function \emph{evaluate}, which returns a set of atoms that belong to the predicate of the head of $r$.
For the evaluation of recursive rules, the algorithm takes advantage of a set $W$, used as a working set to accumulate the atoms of recursive predicates in the evaluation.
The computation of constraints that are returned is done at the end of the bottom-up evaluation (from line~\ref{alg:constr-start}) and takes advantage of the algorithm $\mathit{BuildConstraint}$ described in the following.
Note that for constraints we use the function \emph{ground} which extends \emph{evaluate} to produce 
ground constraints $C$ generated from $\lambda$ w.r.t. $M_{\pi^\prime}$.

\begin{algorithm}[t!]
\small
\caption{BuildConstraint()}\label{alg:buildconstraint}
\begin{algorithmic}[1]
\REQUIRE A constraint $c$, an interpretation $M_{\pi^\prime}$ of $\pi^\prime$, an answer set $M_{ext}$, the program $\lambda_R$
\ENSURE  A ground constraint 
\STATE $R =\emptyset, S = \emptyset$
\WHILE{$c \neq \emptyset$}
\STATE $l = NextLiteral(c)$
\STATE $S = S \cup \{l\}$
\IF{$\cP(l) \in \cP(\pi^\prime)$}
\STATE $R = R \cup \{l^\prime \in M_{\pi^\prime} \mid l^\prime \doteq l\}$
\ELSIF{$\cP(l) \notin \cP(\pi^\prime) \land positive(l)$}
\FORALL{$r \in \lambda_R \mid l \stackrel{\sigma}{=} head_r $}
\FORALL{$b \in body_r $}
\STATE $c = c \cup \{\sigma(b)\}$
\ENDFOR
\ENDFOR
\ELSIF{$\cP(l) \notin \cP(\pi^\prime) \land negative(l)$}
\FORALL{$r \in \lambda_R \mid l \stackrel{\sigma}{=} \naf{head_r} $}
\FORALL{$b \in body_r $}
\STATE $c = c \cup \{\sigma(\overline{b})\}$
\ENDFOR
\ENDFOR
\ENDIF
\STATE  $c = c \setminus S$
\ENDWHILE
\RETURN $toConstraint(R)$
\end{algorithmic}
\end{algorithm}

\paragraph{Handling Failed Constraints.}
We now describe how the constraints to be added to $\pi^\prime$ are computed.
A non-trivial issue is that the constraints in the compiled program might consist of literals that do not appear in $\pi^\prime$.
Algorithm~\ref{alg:buildconstraint} presents a simplified pseudo-code of the procedure that we adopt in our implementation.
The idea is to build a result set $R$ of literals step by step starting from a ground constraint $c$.
Note that $c$ is initially ground, but during the execution of the algorithm non-ground literals might be added to it.
In the following, we use the standard concept of variable-substitution $\sigma$ that represents a mapping from variables to either constants or variables.
At each step, the algorithm selects one literal $l$ in $c$ (function \textit{NextLiteral(c)}).
If the predicate of $l$ appears in $\pi^\prime$ we add all the literals $l'$ in $M_{\pi^\prime}$ that \textit{unifies} (symbol $\doteq$) it, i.e. there is a variable-substitution $\sigma$ such that $\sigma(l) = l'$.
Otherwise, if the predicate of $l$ does not appear in $\pi^\prime$ and $l$ is a positive literal, we add $\sigma(b)$, where $b$ is a body literal of a rule whose head unifies with substitution $\sigma$ (symbol $\stackrel{\sigma}{=}$) with $l$.
Finally, if the predicate of $l$ does not appear in $\pi^\prime$ and $l$ is a negative literal, we add $\sigma(\overline{b})$, where $b$ is a body literal of a rule whose negated head (denoted as $\naf{head_r}$) unifies with substitution $\sigma$ (symbol $\stackrel{\sigma}{=}$) with $l$.
The process continues until $c$ becomes empty.
The set of literals $S$ stores literals that have already been processed to prevent loops.
Note that Algorithm~\ref{alg:buildconstraint} starts from $c$ that is known to be not satisfied in $M_{ext}$, and traces back (like in a top-down evaluation of a query) the computation of $c$ from $\lambda$ to identify a set of literals from $M_{\pi^\prime}$ that imply $c$.
Indeed, steps 7--10 replace a positive literal $l\in c$ by the body of a rule that can infer $l$, whereas steps 11--14 replace a negative literal $\naf l\in c$ with the negation of the body of the rules that could infer $l$ but did not, and 5--6 instantiate the remaining literals in $c$ w.r.t $M_{\pi^\prime}$.
Thus, at the end of the process, $R$ will contain some literals in $M_{\pi^\prime}$ that caused the derivation of $c$ from $\lambda$ and $M_{\pi^\prime}$. Termination is guaranteed, since the same literal is not processed twice (step 15) and steps 7--14 replace literals until no $l$ can be further replaced.


\paragraph{Compilation.} 
The general purpose bottom-up evaluation strategy described above constitutes the template that is instantiated by the compiler depending on the program in input.
In particular, the parts of Algorithm \ref{alg:bottom-up} outlined in blue (i.e. lines 2--7, 10, 12--14, and 20--22) contain instructions that can be evaluated at compile-time because they depend on the syntactic structure of the input;  and thus they are subject to loop unrolling and dead code elimination. Moreover, the dependency graph and its SCCs are computed at compile time and eliminated after unrolling the loops mentioning them.
The parts of the algorithm in black cannot be simplified and are kept in the compiled version to be executed at runtime.
Thus, the compiler given a compilable program $\lambda$ produces an ad-hoc procedure obtained by applying the transformations mentioned above to Algorithm \ref{alg:bottom-up}, and obtains $\lambda^{eval}$ (see Algorithm~\ref{alg:solving}).
Note that the output of the compiler is a procedure that computes the same result of Algorithm \ref{alg:bottom-up} only for the given $\lambda$.

\section{Implementation and Experiments}

The strategy has been implemented within the \wasp solver by exploiting its C++ APIs. The fact that the implementation is embedded into a state-of-the-art ASP solver makes partial compilation more appealing due to the possibility to rely on the consolidated performance of a CDCL solver.
In particular, when the solver starts it calls our implemented compiler, which compiles the input compilable program into a C++ dynamic library that implements a lazy propagator. Candidate models are passed to the dynamic library that computes the extended model and checks the constraints.
The implementation is available at \url{https://bitbucket.org/bernardo_cuteri/lazy_wasp}.

We experimented with partial compilation in four different settings:
\begin{itemize}
\item[] (E1) Compilation of stratified programs;
\item[] (E2) Partial compilation of constraints;
\item[] (E3) Partial compilation of rules and constraints; and 
\item[] (E4) Partial compilation of rules.
\end{itemize}
Time and memory for each run are limited to 10 minutes CPU-time and 6GB, respectively. In all experiments, we compare our system against the best ASP systems for the benchmark at hand.
Concerning experiment (E1), ASP solvers are not included since the programs are already evaluated by ASP grounders.
Concerning experiments (E2), (E3) and (E4), \textsc{clasp} and \textsc{wasp} are used as a reference. Moreover, \textsc{clasp}, \textsc{wasp}, and compilation-based approach use \textsc{gringo} as grounder.
In addition, it should be noted that, being based on \textsc{wasp}, the most relevant result is given by how the compilation-based implementation compares with plain \wasp.

Compilation times are reported exactly once per domain (thus, only on one instance) because the system automatically avoids compiling twice the same program, using an MD5 hash on the compiled program. This fits real-world use-cases where the program is fixed and the instance changes.
In general, compilation times are negligible (up to 2.6 seconds) since we are compiling few rules (up to 15), the only exception being the \emph{wine} encoding in OpenRuleBench that consists of 999 rules and takes some minutes to compile.

For what concernes what parts of the input programs are compiled, we report that in experiment (E1) we compile the whole program, while in all the others we find experimentally some sub-programs that are hard to ground. Sub-programs selection, is in general non trivial, but in many practical cases one can try to incrementally remove parts of the input program, respecting the compilability condition, until grounding becomes acceptable (e.g. the grounding step terminates in an acceptable amount of time).

For all experimental settings, we selected pre-existing benchmarks wherever possible and considered two new benchmarks (\emph{connected k-cut, min-cut with transitive closure}) in the cases where we could not find any. 
New benchmarks consist of classical computer science problems possibly extended to fit the experiment use-case, naively encoded in ASP.

The results are commented in the following in a separate paragraph for each setting. The benchmarks are available for download at \url{https://bitbucket.org/bernardo_cuteri/lazy_wasp}.

\begin{figure}[t!]
  \beginpgfgraphicnamed{orbfig}%
	\begin{tikzpicture}[scale=1]
	\pgfkeys{%
		/pgf/number format/set thousands separator = {}}
	\begin{axis}[
	scale only axis
	, font=\small
	, x label style = {at={(axis description cs:0.5,-0.08)}}
	, y label style = {at={(axis description cs:-0.08,0.5)}}
	, xlabel={\large{Number of instances}}
	, ylabel={\large{Execution time (s)}}
	, xmin=0, xmax=40
	, ymin=0, ymax=610
	, legend style={at={(0.18,0.96)},anchor=north, draw=none,fill=none}
	, legend columns=1
	, width=0.7\textwidth
	, height=0.3\textwidth
	, ytick={0,200,400,600}
	, xtick={0,10,20,30,40}
	, major tick length=2pt
	]
	
	\addplot [mark size=2.5pt, color=black, mark=o] [unbounded coords=jump] table[col sep=semicolon, y index=4] {./ORB.csv}; 
	\addlegendentry{\textsc{compiled}}
	
	\addplot [mark size=2.5pt, color=blue, mark=square] [unbounded coords=jump] table[col sep=semicolon, y index=1] {./ORB.csv}; 
	\addlegendentry{\textsc{gringo}}
	
	\addplot [mark size=2.5pt, color=red, mark=diamond] [unbounded coords=jump] table[col sep=semicolon, y index=2] {./ORB.csv}; 
	\addlegendentry{\textsc{dlv}}

	\addplot [mark size=2.5pt, color=orange, mark=diamond*] [unbounded coords=jump] table[col sep=semicolon, y index=3] {./ORB.csv}; 
	\addlegendentry{\textsc{i-dlv}}	
	\end{axis}
	
	\end{tikzpicture}
  \endpgfgraphicnamed%
	\caption{(E1) OpenRuleBench benchmark}\label{fig:orb}
\end{figure}
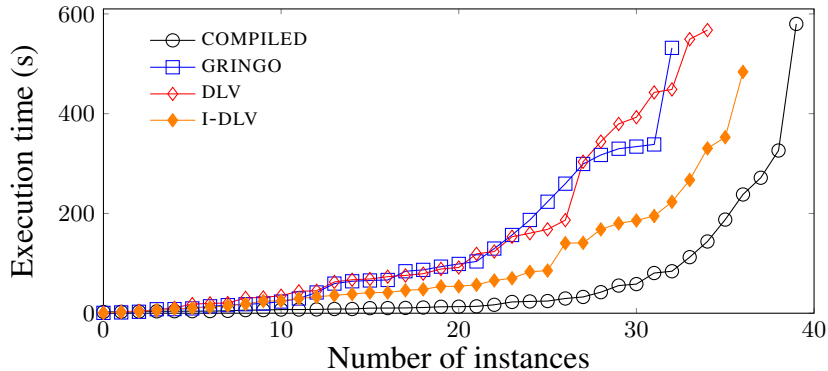

\leanparagraph{(E1) Evaluation of stratified programs.}
Stratified programs are a large subset of ASP programs that allows to model and solve deductive database applications~\cite{DBLP:conf/rweb/EiterIK09}.
To test our implementation, we considered the well-known benchmarks called \textit{OpenRuleBench}, which is an open community benchmark designed to test rule engines. In particular, we run perfect model computation as done for comparing ASP implementations in \cite{DBLP:journals/ia/CalimeriFPZ17}. 
We compared our method with three state-of-the-art ASP systems: \textsc{gringo}~\cite{DBLP:conf/iclp/GebserKKOSW16}, \textsc{dlv}~\cite{DBLP:journals/tocl/LeonePFEGPS06}, and \textsc{i-dlv}~\cite{DBLP:journals/ia/CalimeriFPZ17}.
Plain \wasp is not included in this benchmark since stratified programs are already solved by grounders.
Results are reported in a cactus plot in Figure \ref{fig:orb} and clearly show the performance benefits of the compilation-based approach. Indeed, it solves more instances than state-of-the-art approaches and has in general lower running time.

\begin{table}[b!]
	\caption{(E2) Stable Marriage: Number of solved instances and average running time (in seconds).} \label{tab:stable-compiled}
	\centering
	\footnotesize
	\setlength{\tabcolsep}{0.6em}
	\begin{tabular}{rrrrrrrrrrr}		
		\toprule
		\textbf{Pref. (k\%)} & \multicolumn{2}{c}{\textbf{\textsc{clasp}}} & \multicolumn{2}{c}{\textbf{\wasp}}	& \multicolumn{2}{c}{\textbf{\wasp \textsc{python}}} & \multicolumn{2}{c}{\textbf{\textsc{compiled}}}	&\\
		& \textbf{sol.} & \textbf{avg t} 
		& \textbf{sol.} & \textbf{avg t} & \textbf{sol.} & \textbf{avg t}  & \textbf{sol.} & \textbf{avg t} \\
		0&10&4.36&10&6.2&10&5.8&10&5.6\\
		5&10&28.3&10&25.3&10&5.7&10&5.8\\
		10&10&43.6&8&48.2&10&5.4&10&5.6\\
		15&10&57.9&9&38.3&10&6.8&10&5.6\\
		20&10&62.9&9&50&10&5.9&10&5.4\\
		25&10&67.8&7&52.6&10&5.9&10&5.9\\
		30&10&72.8&10&60.1&10&6&10&5.7\\
		35&10&84.4&5&111.4&10&6.3&10&8.3\\
		40&10&87.6&7&63.3&10&9.4&10&20\\
		45&10&92.0&8&83.8&10&6.3&10&11.3\\
		50&10&94.7&9&67.9&10&6.4&10&8.3\\
		55&10&95.13&7&124.4&9&7.2&9&9.4\\
		60&10&96.36&8&63.3&10&11.5&9&10.7\\
		65&10&99.8&6&66.7&6&18.2&9&17.1\\
		70&10&98.9&6&71&3&21.8&5&132.3\\
		75&10&96.0&8&89.9&0&-&1&13.8\\
		80&10&99.3&7&148.9&0&-&0&-\\
		85&10&107.7&6&107.2&0&-&0&-\\
		90&10&278.7&9&152.2&0&-&0&-\\
		95&8&295.6&10&70.3&0&-&0&-\\
		100&10&98.8&8&61.9&1&7.3&0&-\\
    \midrule
		\textbf{Tot solved}&206& &167& &139& &143& \\
		\bottomrule
	\end{tabular}
\end{table}

\leanparagraph{(E2) Partial compilation of constraints.}
In this experiment, we considered two benchmarks presented in~\cite{DBLP:journals/tplp/CuteriDRS17}, namely StableMarriage and Natural Language Understanding (NLU).
As shown in~\cite{DBLP:journals/tplp/CuteriDRS17}, the encodings of such benchmarks include some constraints leading to a grounding bottleneck. \citeN{DBLP:journals/tplp/CuteriDRS17} presented a strategy to lazily evaluate such constraints by means of custom Python scripts.
Therefore, in the analysis, we compare our approach with these custom Python scripts.

The Stable Marriage benchmark is based on the well-known Stable Marriage problem where there are $n$ men and $m$ women, where each person has a preference order over the opposite sex and the problem consists in finding a marriage that is \emph{stable} (i.e. there is no couple for which both partners would rather be married with each other than their current partner). Results are reported in Table~\ref{tab:stable-compiled}. Each table row is associated to a different value of a parameter $k$ of preferences, e.g. each man (resp. woman) gives the same preference to all the women (resp. men) but to $k\%$ of them a lower preference is given.

The NLU benchmark is about an application of ASP to Natural Language Understanding involving the computation of optimal solutions for
First Order Horn Abduction problems under cost functions cardinality, cohesion, and weighted abduction.
Results are reported in Table~\ref{tab:nlu-compiled} and Figure~\ref{fig:nlu}. Each row in the table presents the result obtained for a specific cost function, while the figure presents the cumulative results for all cost functions.

It is possible to observe that our evaluation strategy works best in the same settings in which constraint lazy instantiators work \cite{DBLP:journals/tplp/CuteriDRS17}, i.e., when the removed constraints are hard to ground, but easy to satisfy. 
The reason is that our evaluation follows the same execution pattern of lazy constraints, i.e., check the constraint on answer set candidates of the original input program without the lazy constraint.
It is important to emphasize here that approaches from \cite{DBLP:journals/tplp/CuteriDRS17} are hand-written by experts, whereas our approach automatically generates the source code with no need of expertise in an imperative language and solver internals/APIs (i.e., the purely declarative solving approach is preserved).

\begin{table}[b!]
	\caption{(E2) NLU Benchmark: Number of solved instances and average running time (in seconds).} \label{tab:nlu-compiled}
	\centering
	\footnotesize
	\setlength{\tabcolsep}{0.6em}
	\begin{tabular}{rrrrrrrrrrr}		
		\toprule
		\textbf{Obj. Func.} & \multicolumn{2}{c}{\textbf{\textsc{clasp}}} & \multicolumn{2}{c}{\textbf{\wasp}}	& \multicolumn{2}{c}{\textbf{\wasp \textsc{python}}} & \multicolumn{2}{c}{\textbf{\textsc{compiled}}}	&\\
		& \textbf{sol.} & \textbf{avg t} 
		& \textbf{sol.} & \textbf{avg t} & \textbf{sol.} & \textbf{avg t} & \textbf{sol.} & \textbf{avg t} \\
card&46&63.7&48&83.0&50&2.8&50&2.3\\
coh&45&68.6&48&83.0&50&26.8&49&18.3\\
wa&46&90.5&48&103.2&49&23.6&49&38.5\\
		\bottomrule
	\end{tabular}
\end{table}
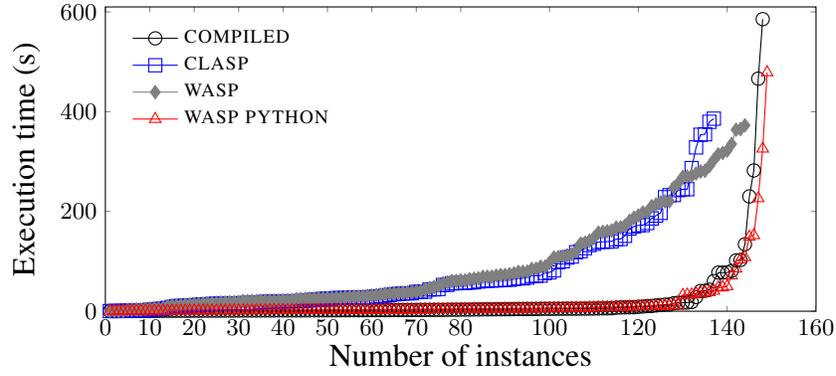
\begin{figure}[t!]
	\beginpgfgraphicnamed{nlu}%
	\begin{tikzpicture}[scale=1]
	\pgfkeys{%
		/pgf/number format/set thousands separator = {}}
	\begin{axis}[
	scale only axis
	, font=\small
	, x label style = {at={(axis description cs:0.5,-0.08)}}
	, y label style = {at={(axis description cs:-0.08,0.5)}}
	, xlabel={\large{Number of instances}}
	, ylabel={\large{Execution time (s)}}
	, xmin=0, xmax=160
	, ymin=0, ymax=610
	, legend style={at={(0.18,0.96)},anchor=north, draw=none,fill=none}
	, legend columns=1
	, width=0.7\textwidth
	, height=0.3\textwidth
	, ytick={0,200,400,600}
	, xtick={0,10,20,30,40,50,60,70,80,100,120,140,160}
	, major tick length=2pt
	]
	
	\addplot [mark size=2.5pt, color=black, mark=o] [unbounded coords=jump] table[col sep=semicolon, y index=3] {./nlp.csv}; 
	\addlegendentry{\textsc{compiled}}
	
	\addplot [mark size=2.5pt, color=blue, mark=square] [unbounded coords=jump] table[col sep=semicolon, y index=1] {./nlp.csv}; 
	\addlegendentry{\textsc{clasp}}
	
	\addplot [mark size=2.5pt, color=gray, mark=diamond*] [unbounded coords=jump] table[col sep=semicolon, y index=2] {./nlp.csv}; 
	\addlegendentry{\textsc{wasp}}	
	
	\addplot [mark size=2.5pt, color=red, mark=triangle] [unbounded coords=jump] table[col sep=semicolon, y index=4] {./nlp.csv}; 
	\addlegendentry{\textsc{wasp} \textsc{python}}	
	\end{axis}
	
	\end{tikzpicture}
	\endpgfgraphicnamed%
	\caption{(E2) NLU Benchmark: Cumulative results of all cost functions.}\label{fig:nlu}
\end{figure}

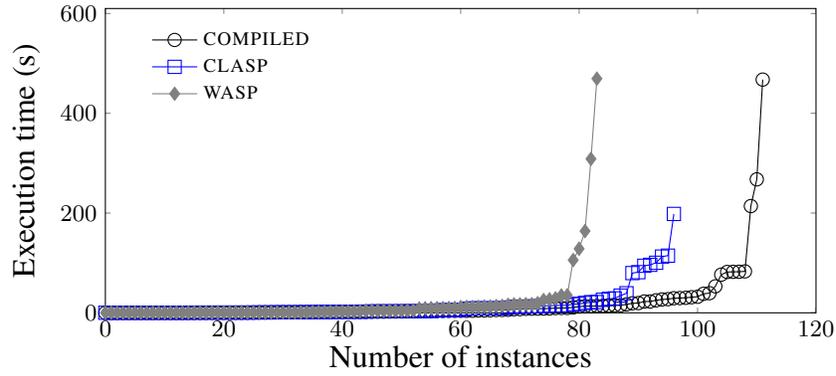
\begin{figure}[t!]
	\beginpgfgraphicnamed{k-cut}%
	\begin{tikzpicture}[scale=1]
	\pgfkeys{%
		/pgf/number format/set thousands separator = {}}
	\begin{axis}[
	scale only axis
	, font=\small
	, x label style = {at={(axis description cs:0.5,-0.08)}}
	, y label style = {at={(axis description cs:-0.08,0.5)}}
	, xlabel={\large{Number of instances}}
	, ylabel={\large{Execution time (s)}}
	, xmin=0, xmax=120
	, ymin=0, ymax=610
	, legend style={at={(0.18,0.96)},anchor=north, draw=none,fill=none}
	, legend columns=1
	, width=0.7\textwidth
	, height=0.3\textwidth
	, ytick={0,200,400,600}
	, xtick={0,20,40,60,80,100,120}
	, major tick length=2pt
	]
	
	\addplot [mark size=2.5pt, color=black, mark=o] [unbounded coords=jump] table[col sep=semicolon, y index=3] {./k_cut.csv}; 
	\addlegendentry{\textsc{compiled}}
	
	\addplot [mark size=2.5pt, color=blue, mark=square] [unbounded coords=jump] table[col sep=semicolon, y index=1] {./k_cut.csv}; 
	\addlegendentry{\textsc{clasp}}
	
	\addplot [mark size=2.5pt, color=gray, mark=diamond*] [unbounded coords=jump] table[col sep=semicolon, y index=2] {./k_cut.csv}; 
	\addlegendentry{\textsc{wasp}}	
	\end{axis}
	
	\end{tikzpicture}
	\endpgfgraphicnamed%
	\caption{(E3) Connected k-cut benchmark}\label{fig:k-cut}
\end{figure}

\begin{figure}[t!]
	\beginpgfgraphicnamed{non-partition}%
	\begin{tikzpicture}[scale=1]
	\pgfkeys{%
		/pgf/number format/set thousands separator = {}}
	\begin{axis}[
	scale only axis
	, font=\small
	, x label style = {at={(axis description cs:0.5,-0.08)}}
	, y label style = {at={(axis description cs:-0.08,0.5)}}
	, xlabel={\large{Number of instances}}
	, ylabel={\large{Execution time (s)}}
	, xmin=0, xmax=120
	, ymin=0, ymax=610
	, legend style={at={(0.18,0.96)},anchor=north, draw=none,fill=none}
	, legend columns=1
	, width=0.7\textwidth
	, height=0.3\textwidth
	, ytick={0,200,400,600}
	, xtick={0,10,20,30,40,50,60,70,80,90,100,110,120}
	, major tick length=2pt
	]
	
	\addplot [mark size=2.5pt, color=black, mark=o] [unbounded coords=jump] table[col sep=semicolon, y index=4] {./non_partition_removal.csv}; 
	\addlegendentry{\textsc{compiled}}
	
	\addplot [mark size=2.5pt, color=blue, mark=square] [unbounded coords=jump] table[col sep=semicolon, y index=2] {./non_partition_removal.csv}; 
	\addlegendentry{\textsc{clasp}}
	
	\addplot [mark size=2.5pt, color=ForestGreen, mark=diamond] [unbounded coords=jump] table[col sep=semicolon, y index=1] {./non_partition_removal.csv}; 
	\addlegendentry{\textsc{alpha}}
	
	\addplot [mark size=2.5pt, color=gray, mark=diamond*] [unbounded coords=jump] table[col sep=semicolon, y index=3] {./non_partition_removal.csv}; 
	\addlegendentry{\textsc{wasp}}	
	\end{axis}
	
	\end{tikzpicture}
	\endpgfgraphicnamed%
	\caption{(E3) Non-partition removal coloring benchmark}\label{fig:non-partition}
\end{figure}
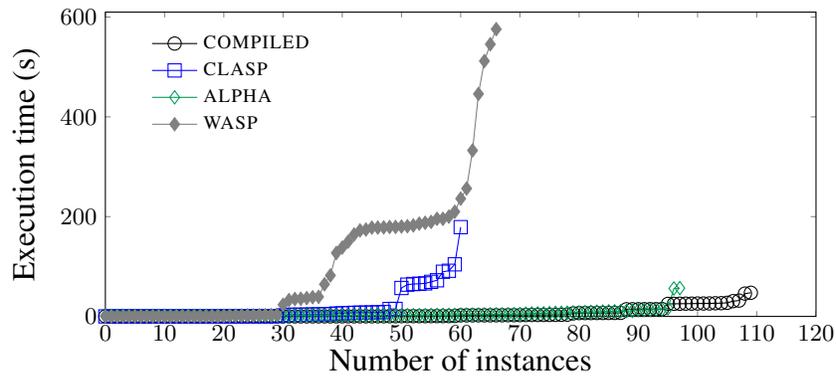
\leanparagraph{(E3) Partial compilation of rules and constraints.}
In this experiment, we consider two benchmarks: 
\emph{connected k-cut} and \emph{non-partition removal coloring}.

Connected k-cut is a graph problem where the goal is to find a cut of size at least \textit{k} such that the two formed partitions are connected. Instances were randomly generated containing graphs with different numbers of nodes (from 200 to 800), different densities (from 0.001 to 0.25) and different cut sizes (from 50 to 800).
Non-partition removal coloring is a benchmark inspired by a real-world configuration application \cite{gebser2015combining} and proposed by \citeN{DBLP:conf/ijcai/BogaertsW18}. The formulation of the problem is as follows: given a directed graph, the goal is to remove one vertex in such a way that the transitive closures of the original and of the resulting graph are equal on the remaining nodes and that the resulting graph is 3-colorable. Instances were taken from~\cite{DBLP:conf/ijcai/BogaertsW18}.

%
Results are reported in Figures~\ref{fig:k-cut} and~\ref{fig:non-partition}.
Concerning \textit{connected k-cut}, compilation-based approach solves 15 and 28 more instances than \textsc{clasp} and \textsc{wasp}, respectively.
Similar results can be observed also in the benchmark \textit{non-partition removal coloring}.
Indeed, compilation-based approach outperforms both \textsc{clasp} and \textsc{wasp}, solving 49 and 43 more instances, respectively.
For the sake of completeness, in this benchmark, we included in the analysis the lazy-solver \textsc{alpha}~\cite{Weinzierl2017}.
Indeed, albeit \textsc{alpha} is not competitive in general with state-of-the-art solvers, in this benchmark it outperforms both \textsc{clasp} and \textsc{wasp}.
However, \textsc{alpha} cannot reach the performance of the compilation-based approach (which solves 12 instances more with similar average running times).

\begin{figure}[t!]
  \beginpgfgraphicnamed{mincut}%
	\begin{tikzpicture}[scale=1]
	\pgfkeys{%
		/pgf/number format/set thousands separator = {}}
	\begin{axis}[
	scale only axis
	, font=\small
	, x label style = {at={(axis description cs:0.5,-0.08)}}
	, y label style = {at={(axis description cs:-0.08,0.5)}}
	, xlabel={\large{Number of instances}}
	, ylabel={\large{Execution time (s)}}
	, xmin=0, xmax=60
	, ymin=0, ymax=610
	, legend style={at={(0.18,0.96)},anchor=north, draw=none,fill=none}
	, legend columns=1
	, width=0.7\textwidth
	, height=0.3\textwidth
	, ytick={0,200,400,600}
	, xtick={0,10,20,30,40,50,60,70,80,100,120}
	, major tick length=2pt
	]
	
	\addplot [mark size=2.5pt, color=black, mark=o] [unbounded coords=jump] table[col sep=semicolon, y index=3] {./mincut_tc.csv}; 
	\addlegendentry{\textsc{compiled}}
	
	\addplot [mark size=2.5pt, color=blue, mark=square] [unbounded coords=jump] table[col sep=semicolon, y index=1] {./mincut_tc.csv}; 
	\addlegendentry{\textsc{clasp}}

	\addplot [mark size=2.5pt, color=gray, mark=diamond*] [unbounded coords=jump] table[col sep=semicolon, y index=2] {./mincut_tc.csv}; 
	\addlegendentry{\textsc{wasp}}	
	\end{axis}
	
	\end{tikzpicture}
  \endpgfgraphicnamed%
	\caption{(E4) Min-cut with transitive closure}\label{fig:mincut}
\end{figure}
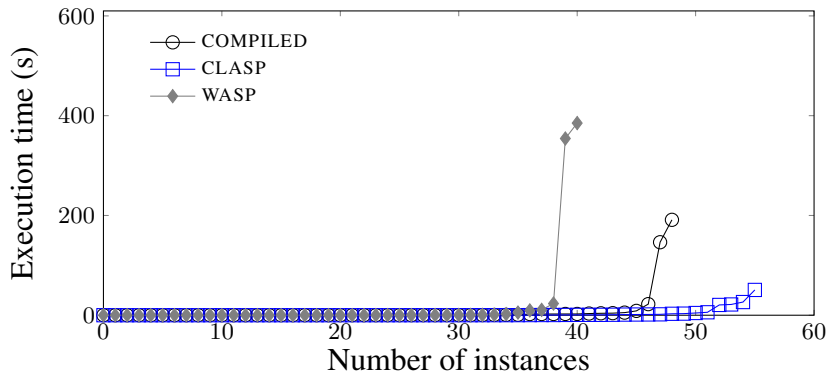

\leanparagraph{(E4) Partial compilation of rules.}
In this experiment, we consider the min-cut problem with transitive closure. Given a graph $G$ the goal is to compute a minimum cost cut of $G$ and to compute the transitive closures of the two resulting partitions.
In order to analyze the performance of compilation-based approach on sub-programs without constraints, in this benchmark the compiled sub-program is only made of rules.
Results are reported in Figure \ref{fig:mincut}, where we observe that \textsc{clasp} is much faster than \textsc{wasp} solving 15 more instances.
Such a gap is partially filled by the compilation-based approach which is able to solve 8 more instances than plain \textsc{wasp}.

\paragraph{Summary of the results.}
Experiments show that the approach is particularly effective for solving stratified programs (E1) and for compiling grounding intensive sub-programs. For what concernes stratified programs, the evaluation is bottom-up as implemented in the other compared systems, but the compilation approach pays off due to its specificity. In experiment (E2), where only constraints are compiled, the approach works similarly w.r.t. the custom lazy instantiators implemented in \cite{DBLP:journals/tplp/CuteriDRS17}: good performances when the constraint is easy to satisfy, but hard do ground. This behaviour has been already shown empirically in \cite{DBLP:journals/tplp/CuteriDRS17} and can easily be observed, for example, in the Stable Marriage results (small values of $k$). In (E3), the approach is effective also in presence of rules. In the \emph{k-cut} benchmark \wasp is originally slower than \clingo, but the compiled approach is faster than \clingo. Moreover, the compiled approach behaves well w.r.t. lazy grounding approaches as shown in the non-partition removal coloring benchmark. Finally, in (E4) the compiled is again able to improve on the performance of the base solver \wasp when only rules (no constraints) are compiled.

\section{Related Work}
Traditional evaluation strategy of ASP systems is based on two steps, namely \textit{grounding} and \textit{solving}; for both phases, several efficient systems have been proposed.
Concerning the grounding, state-of-the-art grounders are \textsc{dlv}~\cite{DBLP:conf/birthday/FaberLP12}, \textsc{gringo}~\cite{DBLP:conf/lpnmr/GebserKKS11} and \textsc{idlv}~\cite{DBLP:journals/ia/CalimeriFPZ17}; which are all based on semi-na\"ive database evaluation techniques~\cite{DBLP:books/cs/Ullman88} for avoiding duplicate work during grounding.
Concerning ASP solvers, the first generation, i.e., \textsc{smodels}~\cite{DBLP:journals/ai/SimonsNS02} and \textsc{dlv} \cite{DBLP:journals/tocl/LeonePFEGPS06}, was based on a DPLL-like algorithm extended with inference rules specific to ASP.
Modern ASP solvers such as \textsc{clasp}~\cite{DBLP:conf/lpnmr/GebserKK0S15} and \wasp~\cite{DBLP:conf/lpnmr/AlvianoDLR15} include mechanisms for conflict-driven clause learning and for non-chronological backtracking. Both solvers also offer an external interface to simplify the integration of custom solving strategies in the main search algorithm. In particular, we used the interface of \wasp to implement the techniques described in the paper.
Alternative approaches are based on the lazy grounding of the whole program, e.g., \textsc{gasp} \cite{DBLP:journals/fuin/PaluDPR09}, \textsc{asperix} \cite{lefevre2017asperix}, or \textsc{alpha} \cite{Weinzierl2017}, where all rules are instantiated lazily; this makes the search less informed but might have a better memory footprint.
These `fully lazy' approaches have in common, that they instantiate even the non-stratified part of the program only when rule bodies of the respective rules are satisfied in the current assignment of the search process, as opposed to our approach where all guesses are instantiated upfront and only stratified parts depending on guesses (including constraints) are computed lazily.
%
Our Algorithm~\ref{alg:buildconstraint} computes constraints
that are related to Justifications \cite{DBLP:conf/ijcai/BogaertsW18},
with the difference that our approach needs ground constraints using only atoms from $\pi^\prime$,
while the \textsc{alpha} solver uses nonground constraints computed from Justifications branches that are cut off at the first negated literal.
CASP~\cite{DBLP:journals/tplp/BalducciniL17,DBLP:journals/tplp/OstrowskiS12} and ASPMT~\cite{DBLP:conf/jelia/BartholomewL14} can solve problems with large constraints, but extend the language with external theories.
The compilable program definition is related to Rule Splitting Sets of HEX programs~\cite{DBLP:journals/tplp/EiterFIKRS16}, 
however, we here define them on the basis of predicates, not partially ground atoms.
ASP Modules \cite{Janhunen2009} are more permissive than compilable subprograms
because they permit mutually cyclic (negative) dependencies among modules,
which is not possible in compilable subprograms.

\section{Conclusion}
Compilation-based approaches are meant to speed up computation by exploiting information known at compilation time to create custom procedures that are specific to the problem at hand.
In this paper, we presented what is, to the best of our knowledge, the first work on compilation-based techniques for ASP solving. In our approach, we allow compilation of ASP sub-programs and we define what a compilable sub-program is, i.e., we specify what are the conditions under which our approach can be adopted. 
The presented approach has been developed as a solver extension of \wasp which is a state-of-the-art ASP solver. 
The evaluation strategy presented includes a bottom-up evaluation for computing the unique stable model of the compilable sub-program and a top-down evaluation for computing failed constraints in terms of literals that are known to the ASP solver.
An experimental analysis shows the benefits that can be obtained in different use-cases by a compilation-based approach. The approach is particularly suited for solving stratified programs, and for compiling ground-intensive sub-programs where lazy instantiators are effective.

In the future, we are planning to extend the presented approach to allow eager/post propagation, i.e., the evaluation is performed also on partial interpretations every time a new literal is chosen (eager) or when unit propagation ends (post). 
Moreover, it is also interesting to investigate whether it is possible to automatically select a sub-program to be compiled that maximizes the performance of our technique.

\section*{Acknowledgments}
This work has been partially supported by MIUR under PRIN 2017 project n. 2017M9C25L 001 (CUP H24I17000080001),
and from the EU's Horizon 2020 research and innovation program under grant agreement No 825619 (AI4EU).


\bibliographystyle{acmtrans}
\bibliography{refs}

\label{lastpage}
\end{document}